\documentclass[traditabstract,longauth]{aa}
\usepackage{times}
\usepackage{natbib}
\usepackage[dvips]{graphicx}
\bibpunct{(}{)}{;}{a}{}{,}

\def\grb{GRB 080514B}
\newcommand{\degree}{^{\circ}}

\def\ltsima{$\; \buildrel < \over \sim \;$}
\def\lsim{\lower.5ex\hbox{\ltsima}}
\def\gtsima{$\; \buildrel > \over \sim \;$}
\def\gsim{\lower.5ex\hbox{\gtsima}}

\begin{document}

\title{AGILE detection of \\delayed gamma-ray emission from GRB 080514B}

 \author{
  A.~Giuliani\inst{1},
  S.~Mereghetti\inst{1},
  F.~Fornari\inst{1},
  E.~Del Monte\inst{3},
  M.~Feroci\inst{3},
  M.~Marisaldi\inst{5},
  P.~Esposito\inst{1,7,18},
  F.~Perotti\inst{1},
  M.~Tavani\inst{3,4},
   A.~Argan\inst{3},
   G.~Barbiellini\inst{6},
   F.~Boffelli\inst{7,18},
   A.~Bulgarelli\inst{5},
   P.~Caraveo\inst{1},
   P.~W.~Cattaneo\inst{7},
   A.W.~Chen\inst{1,2},
   E.~Costa\inst{3},
   F.~D'Ammando\inst{3,4},
  G.~Di Cocco\inst{5},
  I.~Donnarumma\inst{3},
  Y.~Evangelista\inst{3},
   M.~Fiorini\inst{1},
  F.~Fuschino\inst{5},
   M.~Galli\inst{14},  
  F.~Gianotti\inst{5},
  C.~Labanti\inst{5},
   I.~Lapshov\inst{3},
  F.~Lazzarotto\inst{3},
  P.~Lipari\inst{12},
 F.~Longo\inst{6},
  A.~Morselli\inst{9},
  L.~Pacciani\inst{3},
   A.~Pellizzoni\inst{1},  
    G.~Piano\inst{3,4},
   P.~Picozza\inst{9},
   M.~Prest\inst{10},  
   G.~Pucella\inst{3},
   M.~Rapisarda\inst{3},
   A.~Rappoldi\inst{7},
  P.~Soffitta\inst{3},
 M.~Trifoglio\inst{5}, 
  A.~Trois\inst{3},
 E.~Vallazza\inst{6},
 S.~Vercellone\inst{1},  
 D.~Zanello\inst{12},
 L.~Salotti\inst{13},
S.~Cutini\inst{8}, C.~Pittori\inst{8}, B.~Preger\inst{8}, P.~Santolamazza\inst{8}, F.~Verrecchia\inst{8},
N.~Gehrels\inst{15}, K.~Page\inst{16}, D.~Burrows\inst{17},
A.~Rossi\inst{19},
K. ~Hurley\inst{20}, I.~Mitrofanov\inst{21}, W.~Boynton\inst{22}
}

\institute{$^1$INAF/IASF--Milano, Via E.~Bassini 15, I-20133 Milano, Italy \\
$^2$CIFS--Torino, Viale Settimio Severo 63, I-10133 Torino, Italy \\
$^3$INAF/IASF--Roma, Via del Fosso del Cavaliere 100,  I-00133 Roma, Italy \\
$^4$Dip. di Fisica, Univ. ``Tor Vergata'', Via della Ricerca  Scientifica 1, I-00133 Roma, Italy \\
$^5$INAF/IASF--Bologna, Via Gobetti 101, I-40129 Bologna, Italy \\
$^6$Dip. di Fisica and INFN Trieste, Via Valerio 2, I-34127 Trieste, Italy\\
$^7$INFN--Pavia, Via Bassi 6, I-27100 Pavia, Italy\\
$^{8}$ASI--ASDC, Via G. Galilei, I-00044 Frascati (Roma), Italy\\
$^{9}$INFN--Roma ``Tor Vergata'', Via della Ricerca Scientifica 1,  I-00133 Roma, Italy\\
$^{10}$Dip. di Fisica, Univ. dell'Insubria, Via Valleggio 11,  I-22100 Como, Italy\\
$^{11}$ENEA--Roma, Via E. Fermi 45, I-00044 Frascati (Roma), Italy\\
$^{12}$INFN--Roma ``La Sapienza'', Piazzale A. Moro 2, I-00185 Roma,  Italy\\
$^{13}$ASI, Viale Liegi 26 , I-00198 Roma, Italy\\
$^{14}$ENEA, Via Martiri di Monte Sole 4, I-40129 Bologna,  Italy\\
$^{15}$NASA Goddard Space Flight Center, Greenbelt, MD 20771, USA\\
$^{16}$Department of Physics and Astronomy, University of Leicester, Leicester LE1 7RH, UK\\
$^{17}$Department of Astronomy and Astrophysics, Pennsylvania State University, University Park, PA 16802, USA\\
$^{18}$Dip. di Fisica Nucleare e Teorica, Univ. degli Studi di Pavia, Via A. Bassi 6, I-27100, Pavia, Italy\\
$^{19}$Th\"uringer Landessternwarte Tautenburg, Sternwarte 5, D-07778 Tautenburg, Germany\\
$^{20}$University of California, Berkeley, Space Sciences Lab, 7 Gauss Way, Berkeley, CA 94720, USA\\
$^{21}$Institute for Space Research, Profsojuznaja 84/32, Moscow 117997, Russia\\
$^{22}$Lunar and Planetary Laboratory, University of Arizona, Tucson, AZ 85721, USA}

\offprints{A. Giuliani, giuliani@iasf-milano.inaf.it}

\date{Received / Accepted}

\authorrunning{A. Giuliani et al.}
\titlerunning{\textit{AGILE} \grb\ }

\abstract{ \grb\   is the first  {gamma ray burst (GRB)}, since the time of EGRET,
for which individual photons of energy above several tens of MeV have been
detected with a pair-conversion tracker telescope. This burst was
discovered with the Italian \textit{AGILE} gamma-ray satellite. 
The GRB was localized with a cooperation by \textit{AGILE} and  {the interplanetary network (IPN)}. 
The gamma-ray imager (GRID) estimate of the position, obtained before the SuperAGILE-IPN localization, is 
found to be consistent with the burst position.
{The hard X-ray emission observed by SuperAGILE lasted about 7 s, 
         while there is evidence that the emission above 30 MeV extends 
         for a longer duration (at least ~13 s).}
Similar behavior was seen in the past from
a few other GRBs observed with EGRET. 
However, the latter
measurements were affected, during the brightest phases, by
instrumental dead time effects, resulting in only lower limits to
the burst intensity. 
Thanks to the small dead time of the \textit{AGILE}/GRID we could assess that in the case of \grb\ the gamma-ray to
X--ray flux ratio changes significantly between the prompt and
extended emission phase.
 }

\maketitle

\section{Introduction}

Only a relatively small number of gamma-ray bursts (GRBs) have
been detected so far at energies above the MeV region.  The
largest sample of high-energy observations has been collected
using the large calorimeter of the EGRET instrument on board the  {Compton Gamma Ray Observatory} (\textit{CGRO}).
This detector provided GRB
light curves and spectra in the 1--200 MeV energy range, thus
extending the spectral coverage of the brightest bursts discovered
at lower energy with the  {Burst And Transient Source Experiment} 
\citep[BATSE][]{kaneko2008}. Only for a handful
of GRBs were high-energy photons detected also with the EGRET
spark chamber, {sensitive in the energy range 30 MeV - 10 GeV} 
\citep[see, e.g.][and references therein]{dingus2001}.
These observations showed the existence of a few bursts with very
hard spectra, with peak energy up to \gsim170 MeV,  and possibly
of GRBs with  high energy ``excesses'', {i.e. an emission above 100 MeV larger than the extrapolation of their X-rays spectra}. 
It is particularly
important for theoretical models to understand whether the
high-energy emission is a separate spectral component, possibly
with a different time evolution from than of the prompt emission. This
possibility is supported by the slow time decay of the high-energy
flux in GRB 941017 \cite{gonzalez2003}, as well as by the EGRET
detection of delayed photons from GRB 940217 \cite{hurley1994},
including one with an energy of 18 GeV after 1.3 hours.
Two factors that affected the study of GRBs at high energy are
intrinsic to gamma-ray imagers of the old generation, based on
spark chamber trackers: the relatively small field of view,
limiting the sample of observed GRBs,  and the significant
instrumental dead time, reducing the number of detected photons in
the brightest parts of the bursts, thus preventing a reliable
measure of their peak flux and fluence.

Both limitations are overcome by new gamma-ray satellites using
self-triggering trackers based on silicon microstrip
technology, like \textit{AGILE} \cite{tavani}. The \textit{AGILE} Gamma-Ray Imaging
Detector (GRID), operating in the  30 MeV -- 50 GeV energy range,
has a field of view as large as one fifth of the whole sky and a
dead time of only $\sim$200 $\mu$s, which makes it particularly
suited for the observation of GRBs. 
{For comparison, EGRET had a field of view
of $\sim$1 sr and a dead time of $\sim$200 ms}. 
Furthermore, \textit{AGILE} carries other detectors that allow the study of
GRB emission in different energy ranges: SuperAGILE provides two
one-dimensional images (along orthogonal directions), light
curves and spectra in the 17--50 keV range \cite{Feroci2007}. The
MiniCalorimeter (MCAL) \cite{Labanti2008}, besides being used as
part of the GRID, can be used to autonomously detect and study
GRBs in the 0.35--100 MeV range. 
Finally, GRB light curves in the hard X-ray band can be obtained from the GRID Anti
Coincidence scintillator panels \citep[ACS,][]{perotti}. 
The SuperAGILE capabilities in terms of rapid and accurate localizations 
of GRBs have been well demonstrated already with the discovery of its 
first GRB, 070724B \citep{Del_Monte_et_al_2008} and continue with a rate of
about 1 GRB per 1--2 months. Instead, MCAL detects approximately 1 GRB per week,
as described in \citep{Marisaldi2008}.

Here we present the observation of the first GRB for which a
positive detection has been obtained by the GRID. 
Interestingly,
the AGILE data give strong evidence that the GRB duration at
energies above 30 MeV is at least double than that observed at
lower energies, suggesting that different emission processes might
be at work in the gamma and X-ray range.
\begin{figure}[ht!]%
\centering
\includegraphics[angle=90, width=9. cm]{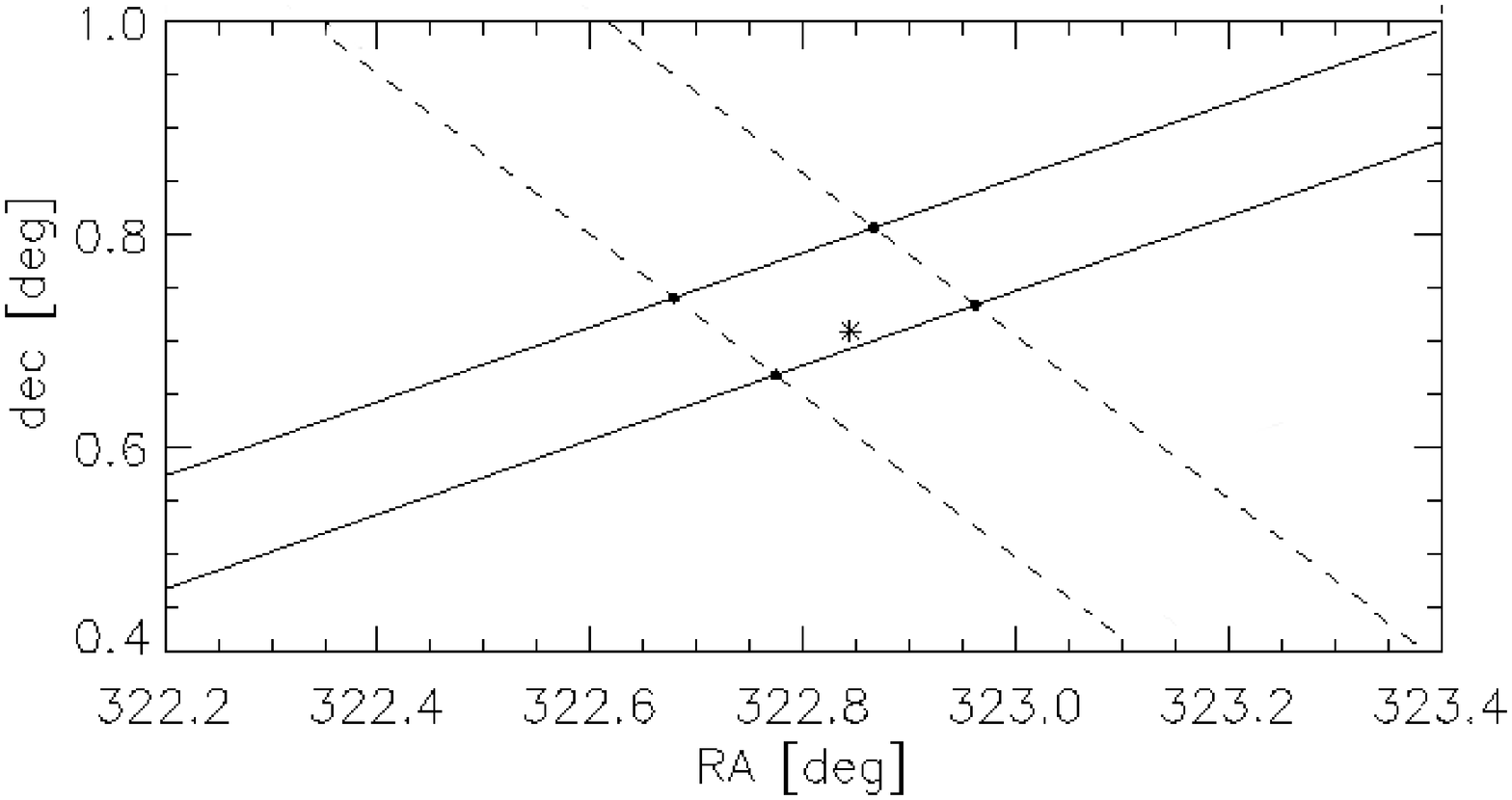}
\caption{
Combined SuperAGILE and IPN localization of GRB 080514B.
The star ($*$) marks the position of the afterglow 
found in X-rays by the Swift-XRT \citep[see][]{Page08_GCN7723} and in optical by IAB80 telescope  
\citep[see][]{GCN_7719, GCN_7720}, 
the solid lines indicate the SuperAGILE error box and the dashed lines represent the
IPN triangulation based on SuperAGILE and Mars Odyssey data. 
}
\label{fig:SuperAGILE-IPN_position}
 
\vspace{-0.4 cm}
 
\end{figure}

\section{Observations and data analysis}

\subsection{SuperAGILE}

SuperAGILE detected \grb\ on 2008 May 14  at 09:55:56 UT (hereafter $t_0$).
The GRB triggered the on-board  detection system
\citep{Del_Monte_et_al_2007} and was localized  within
the 1-D region of the SuperAGILE field of view. The detection was
confirmed by the on-ground trigger, yielding an average signal-to-noise 
ratio of 15.2 on the 4.096 s timescale. The burst was at an
off-axis angle of $-37.61\degree$
along the X coding direction, and outside the field of view in the
Z direction.
This resulted in an error box consisting of a long and narrow (15
arcmin total width) rectangle centered at $\mathrm{R.A.=322.104 \degree\ }$, 
$\mathrm{Dec.=0.486 \degree\ }$
(J2000), and extending toward $\mathrm{R.A. = 302.9 \degree}$,
$\mathrm{Dec. = -6.2 \degree}$ and $\mathrm{R.A. = 341.3\degree}$,  $\mathrm{Dec. = +7.1 \degree}$.
\grb\ was also detected by \textit{Mars Odyssey}, \textit{MESSENGER}, Konus-\textit{Wind}, \textit{Suzaku}/ {Wideband All-sky Monitor (WAM)}, and \textit{INTEGRAL}/SPI- {Anticoincidence Shield} in the Interplanetary Network for GRB localization.  Initially, an Odyssey-SuperAgile annulus was obtained \cite{GCN_7715}. 
{Combining the SuperAGILE
error box with the IPN annulus gives the error region, with an area of about 100 square arcmin, 
shown in  Fig. \ref{fig:SuperAGILE-IPN_position}. }

{Two observations of the afterglow of \grb\ with the \textit{Swift} X-Ray Telescope \citep[XRT;][]{burrows05} were obtained as ToO.
	 A bright fading source was detected inside the SuperAGILE-IPN error box
         \citep{Page08_GCN7723}
	 An optical afterglow was discovered  by observing the 
         complete IPN error box \cite{GCN_7719}, 
         and the probable host galaxy was identified by Perley et al. (GCN 7874), 
         who constrained the redshift to be $<$ 2.3 .
	 } 

\begin{figure}
 \centering
 \includegraphics[angle=00, width=9.5 cm, height=12. cm]{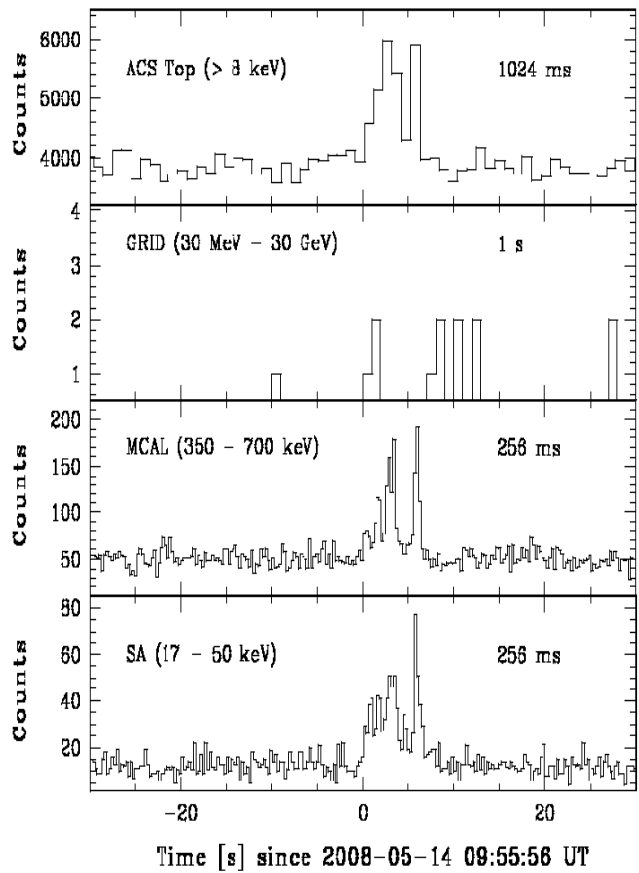}
 \caption{Lightcurves of \grb\ as measured with the different detectors on
 board the \textit{AGILE} satellite.
 The SuperAGILE light curve has been obtained  selecting
 only the illuminated region of the detector.} 
 \label{fig:multi_lc}
 
 \vspace{-.5 cm}
 
 \end{figure}

The SuperAGILE lightcurve of \grb\ in the 17 -- 50 keV energy band
is shown in the bottom panel of Fig. \ref{fig:multi_lc}.
The burst has a duration of $\sim$ 7 s and a multipeak structure.
From the SuperAGILE imaging analysis,  under the assumption of a
Crab-like spectrum,  we can derive an average flux of 
$\mathrm{2.3 \cdot 10^{-7} \; erg \; cm^{-2} \; s^{-1}}$, a fluence of
$\mathrm{1.6 \cdot 10^{-6} \; erg \; cm^{-2}}$, and a 256 ms peak
flux of $\mathrm{7.3 \cdot 10^{-7} \; erg \; cm^{-2} \; s^{-1}}$
(all the values refer to the 17--50 keV energy band). 
The spectra, for the time interval from
$t_0$ to $t_0+7$ s, were accumulated separately in the two
illuminated detectors and fitted simultaneously. A power law
adequately describes the data in the 20--50 keV energy band, with
a photon index of $-0.02^{+0.47}_{-0.48}$ (at 90 \% confidence
level, with $\chi^2_r$=0.85 for 23 degrees of freedom). Based on
the best fit model, the average flux and the fluence are in
reasonable agreement with the values extracted from the images and
are in good agreement with the extrapolation of the Konus-Wind
values \citep{GCN_7751}.
\subsection{MCAL detection}

\grb\ triggered also the \textit{AGILE}/MCAL detector.  A valid trigger was
issued by the onboard trigger logic at $t_0$, after a simultaneous
partial trigger was detected, in the 1.024~s time window, by two
of the nine detector ratemeters \citep[see][for a detailed description of the MCAL burst trigger logic]{Fuschino2008}.
MCAL data have been collected in photon-by-photon mode for the
time interval between 100~s before and 40~s after the trigger
time, yielding  energy information and a timetag with
$2~\mathrm{\mu s}$ accuracy for each event. 

The MCAL lightcurve (Fig.~\ref{fig:multi_lc}) shows a
multi-peaked structure and a total duration (T90) of about 5.6~s.
According to preliminary in-flight calibration parameters, the time-integrated  spectrum between $t_0$ and $t_0+7~s$ can be fitted with a power law with photon index $2.6 \pm 0.2$. 
The fluence in the 
500-5000~keV energy range is $(7.8 \pm 1.5) \cdot 10^{-6}~\mathrm{erg \; cm^{-2}}$, 
and the 256~ms peak flux, measured at $t_0+6 s$, is $(3.6 \pm 1.1) \cdot 10^{-6}~\mathrm{erg\; cm^{-2}\; s}$ in the same energy band.
No significant spectral evolution at energies higher than 350~keV is observed. 
We note that the MCAL photon index is in good agreement with the high energy photon index 
obtained%
with Konus-\textit{Wind} \cite{GCN_7751} and \textit{Suzaku}/WAM \cite{Hanabata2008}. The observed fluence is also in good agreement with the \textit{Suzaku}/WAM results, while it is about 60$\%$ that obtained by Konus-\textit{Wind}.
{The MCAL GRB detection capabilities and a discussion on the preliminary
in-flight calibration is reported in Marisaldi et al. (2008).}
\begin{figure}
 \centering
 \includegraphics[angle=00, width=9 cm, height=9. cm]{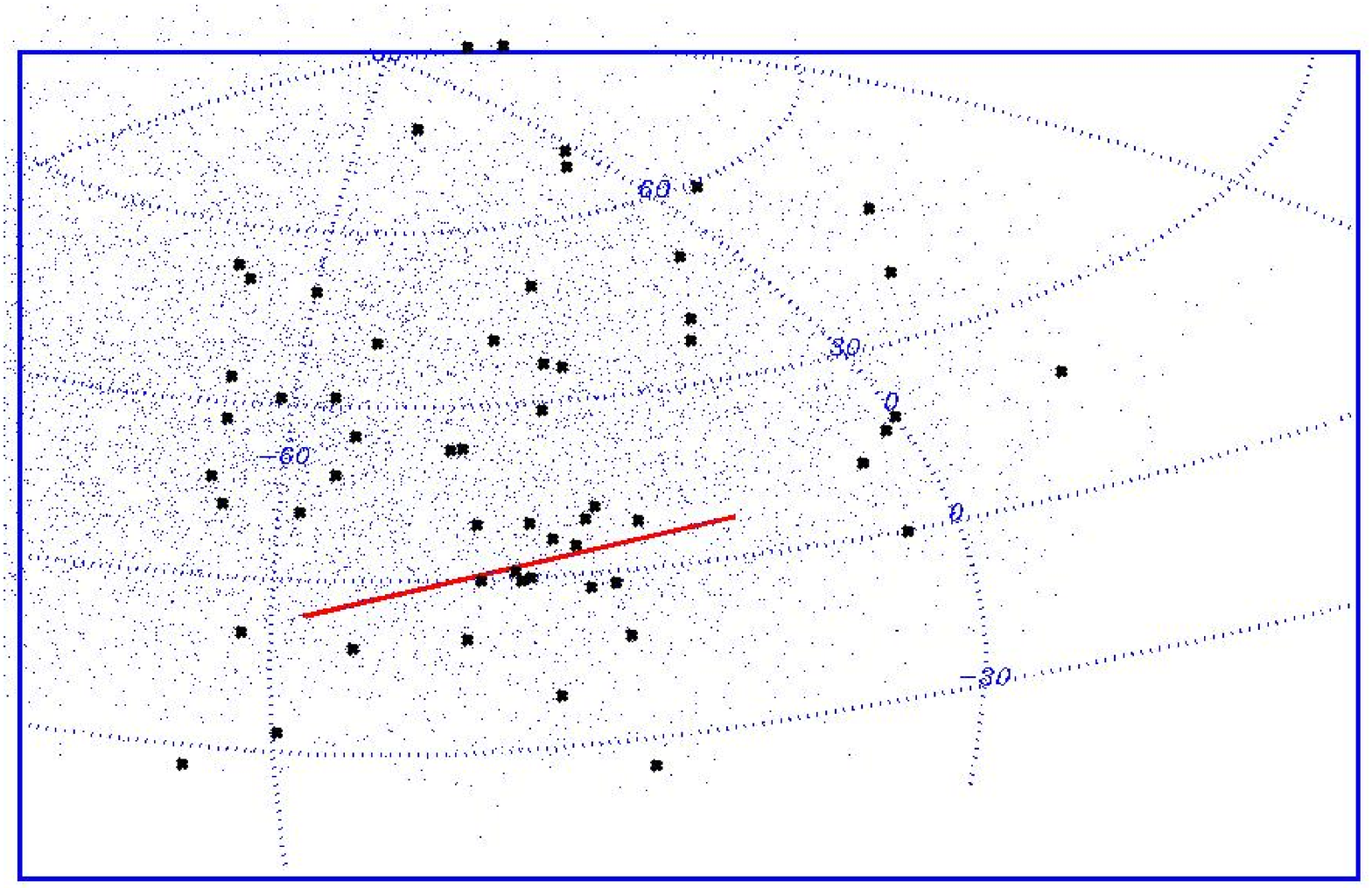}
 \caption{Distribution in  {Celestial} coordinates of all the GRID photons detected
 during the orbit in which \grb\ was detected ($\sim$ 3000 sec exposure time). The
 photons detected from $t_0$ to $t_0$+20 s are marked by stars. The red line
 indicates the SuperAGILE error box. The absence of star marks to the left
 of l $\sim$--70 is due to the Earth occultation during the GRB.  } 
 \label{fig:map}
 
 \vspace{-0.2 cm}
 
 \end{figure}
\begin{figure}
 \centering
 \includegraphics[angle=00, width=9. cm, height=5.6 cm]{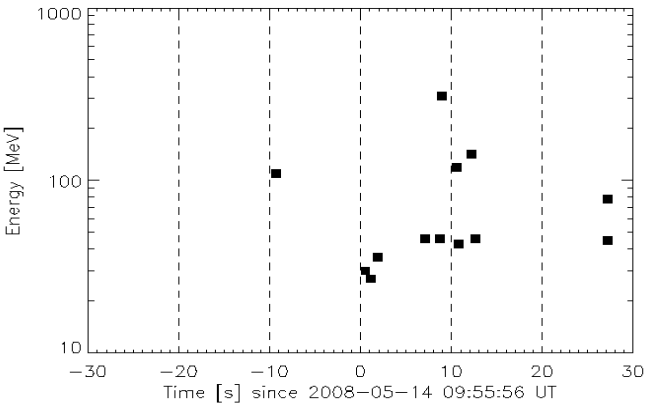}
 \caption{Energy of the photons detected by the GRID from
 the direction of \grb\ }
 \label{fig:EvsT}
 
 \vspace{-0.4 cm}
 
 \end{figure}
\subsection{GRID: high energy emission}

The \textit{AGILE}/GRID provides directional, timing and energy information
for all the detected events. A series of cuts, applied first
on-board (both through hardware-coded and software algorithms) and
subsequently in the ground analysis, allows a discrimination between
celestial gamma-ray photons and particle-induced or Earth albedo
background. In this analysis we used the version F4 of the
Standard Analysis filter which first rejects events that can be
confidently attributed to background particles and then estimates,
through a Kalman filter technique, the energies and arrival directions of
the events classified as photons \cite{Giuliani2006:kalmex}.
To search for gamma-ray emission associated with \grb\, we used
all the events classified as photons with  energy $>$25 MeV.
Taking into account the off-axis position and assuming a power-law
spectrum with photon index $2$, the  effective area corresponding
to this event selection is about 400 cm$^2$ .

Our first search, which led to the preliminary report of a possible
detection \citep{GCN_7716}, was based on the photons collected in
4 time intervals of different durations (5, 10, 20 and 30 s)
starting at $t_0$ within a radius of 15 degrees centered at the
middle of the error region provided by SuperAGILE
($\mathrm{R.A.=322.15 \degree \ } , \mathrm{Dec.=0.5 \degree \ }$). 
This extraction radius corresponds to the 68\% containment
region for the above energy selection and a ``canonical'' power
law spectrum with photon index $\Gamma$=2. To estimate the number
of expected background photons, we used a large annular region and
time intervals before and after the GRB in the same \textit{AGILE} orbit.
This yielded an expected rate of 0.233 counts s$^{-1}$ in the GRB
extraction radius\footnote{this value of the expected background 
rate has been confirmed by the subsequent analysis of longer time
spans of data from other \textit{AGILE} orbits and with the same sky
pointing direction}. Thus the 8 photons observed in the 10 s long
time interval correspond to a chance probability of about $2.8
\cdot 10^{-3}$ of being due to the background. The number of
photons observed in the 5, 20 and 30 s long intervals are
respectively 4, 16 and 20, and correspond to chance probabilities
of $0.031$, $3.1 \cdot 10^{-5}$  and $4.4 \cdot 10^{-5}$.

Independent evidence supporting the gamma-ray detection of \grb\
came from the quick look imaging analysis performed immediately
after the SuperAGILE localization. 
This is illustrated in the sky image shown in Fig.~\ref{fig:map}.  
The black dots indicate the positions of the photons detected during the whole orbit, while
the stars are only those in the 20 s time interval starting at $t_0$. 
{The clustering of the photons (about $2 \cdot 10^{-2}$ $ph$ $deg^{-2}$ in the first 20 seconds) near the center of the long SuperAGILE
strip strongly suggested that the burst was indeed located, as later confirmed by the IPN localization.}

In order to perform a more refined analysis to derive the GRB
spectrum and light curve, we adopt here an energy-dependent
extraction radius based on the instrument 
point spread function\footnote{The background rate corresponding to
this data selection is 0.147 counts s$^{-1}$, so the
statistical significance of the GRID detection is similar to that
derived from the quick look analysis reported above}. 
This leads to the selection of events most likely to be associated with the GRB.
Their energies and times of arrival are plotted in
Fig.~\ref{fig:EvsT} while the 
resulting GRID light curve  is shown in Fig.~\ref{fig:multi_lc}.
A comparison with the data obtained at lower energies immediately
shows that the emission of high energy photons lasted
significantly longer. 
{To compare the high-energy light curve, as inferred from the
sparse GRID photons, with that observed at lower energy, 
we estimated whether the time of arrivals of the 10 GRID photons are consistent
with the distribution expected from the shape of the MCAL and SuperAGILE 
light curves. Using a Kolmogorov-Smirnov test, we  obtained
a probability for this hypothesis of only about 0.003,  confirming that the 
GRB duration is longer in the GRID energy range.
The time of arrivals of the GRID events are instead statistically
consistent with a constant flux over the whole time interval from t$_0$ to t$_0$+20 s, 
contrary to what is seen in all the data at lower energy. }
In this respect it is important to note that the number of
GRID photons in the first time interval, where a brighter emission
could be expected, is not limited by dead time effects.

The distribution of the energies of the GRID photons is
consistent with a power law spectrum of photon index between $3.5$ and $2$, 
in agreement with the value obtained extrapolating the Konus-Wind spectrum ($2.5$).
Assuming a photon index of $2.5$,
{the 25 MeV - 30 GeV fluence in the time 
interval from $t_0$ to $t_0$ + 7 s is (1.4$\pm$0.7)$\cdot$10$^{-2}$ ph cm$^{-2}$, while the fluence  from $t_0$ to $t_0$ + 20 s is (3.3$\pm$1.1)$\cdot$10$^{-2}$ ph cm$^{-2}$.
Due to the large statistical errors affecting these values, both are consistent with the
extrapolation of the spectrum measured at lower energies.}

\section{Discussion}

We have reported evidence for high-energy ($>$25 MeV) emission
from \grb\ , a burst belonging to the class of long duration GRBs,
discovered thanks to the complement of detectors on
board the \textit{AGILE} satellite. The highest-energy photon consistent
with the direction of the burst has an energy of about 300 MeV.
Most of the detected GRB photons are at lower energy, in the range
$\sim$25-50 MeV,
consistent with a power law spectrum with photon index $2.5$.

Previous detections of GRBs
at these high energies were obtained more than 10 years ago with
EGRET, before the discovery of GRB afterglows confirmed their
cosmological origin.
{X-ray and optical/NIR observations indicate that the afterglow properties of \grb\ 
are similar to those of other bursts \cite{rossi2008}}.

The most striking feature of \grb\ concerns the fact that the
arrival times of the high energy photons detected with the \textit{AGILE}/GRID 
do not coincide with the brightest peaks seen at hard X-rays.
Three photons are concentrated within 2 s at the beginning of the
burst, while the next ones arrive only when the  X-ray emission
has returned to a level consistent with the background 
(7 s after the beginning of the burst). 
It is important to note that the number of photons in the initial 2 s 
is not limited by instrumental dead time effects, as was the case for some of the EGRET GRBs
for which only a flux lower limit could be measured in correspondence 
of the brightest peaks.
This implies a rapid time evolution of the gamma-ray to X-ray flux
ratio, although a quantitative assessment of this variability is
hampered by the small statistics. 

{Previous EGRET observations indicated a similar behavior 
for the bursts GRB 941017 \cite{gonzalez2003} and GRB 940217 \cite{hurley1994},
suggesting the possibility that the high-energy emission, at least in some
cases, is not a simple extension of the main component, but originates
from a different emission mechanism and/or region.
Theoretical models predict that Inverse Compton (IC) plays a
significant role in the high-energy emission from GRBs.%
The ratio of IC ($\sim$MeV-GeV) to synchrotron
($\sim$keV) fluences depends on the energy of the relativistic
electrons accelerated in the shocks and on the ratio between
magnetic field and photon densities. 
Detailed time resolved spectroscopy covering the
whole energy range, as now possible with further \textit{AGILE} and \textit{GLAST} observations 
(complemented by low energy data from one
of the X and soft gamma-ray experiments currently on orbit) 
is required to disentangle the different components 
and obtain some constraints on the physical parameters of the sources.}

\begin{acknowledgements}

The \textit{AGILE} Mission is funded by the Italian Space Agency (ASI) with
scientific and programmatic participation by the Italian Institute
of Astrophysics (INAF) and the Italian Institute of Nuclear
Physics (INFN). 
KLP acknowledges financial support from STFC.
KH is grateful for support under the Mars Odyssey Participating Scientist program, JPL grant 1282043.

\end{acknowledgements}
\bibliographystyle{aa}

\end{document}